\documentclass[11pt]{article}

\usepackage{axodraw}
\usepackage{epsfig}

\begin{document}
\setlength{\unitlength}{1mm}

\begin{titlepage}

\begin{flushright}
LAPTH-864/01\\
LPT-Orsay 01-77\\
July 2001
\end{flushright}
\vspace{1.cm}

\begin{center}
\large\bf
{\LARGE\bf Is a Large Intrinsic $k_T$ Needed to Describe\\ 
Photon + Jet Photoproduction at HERA?}\\[2cm]
\rm
{M.~Fontannaz$^{a}$, J.~Ph.~Guillet$^{b}$, G.~Heinrich$^{a}$ }\\[.5cm]

{\em $^{a}$Laboratoire de Physique Th\'eorique\footnote{Unit\'e 
Mixte de Recherche 8627 du CNRS} LPT,\\ 
           Universit\'e de Paris XI, B\^atiment 210,\\
           F-91405 Orsay, France} 

\medskip

{\em $^{b}$Laboratoire d'Annecy-Le-Vieux de Physique 
 Th\'eorique\footnote{Unit\'e Mixte de Recherche 5108 du CNRS, associ\'ee \`a 
              l'Universit\'e de Savoie.} LAPTH,}\\
      {\em Chemin de Bellevue, B.P. 110, \\ 
         F-74941 Annecy-le-Vieux, France}\\[3.cm]

\end{center}
\normalsize

\begin{abstract}
We study the  photoproduction of an isolated photon and a jet 
based on a code of partonic event generator type which includes 
the full set of next-to-leading order corrections. 
We compare our results to a recent ZEUS analysis 
in which an effective $k_T$ of the incoming partons 
has been determined. We find that no 
additional intrinsic $k_T$ is needed to describe the data.

\end{abstract}

\vspace{3cm}

\end{titlepage}

\section{Introduction}

Inclusive photoproduction of an isolated prompt photon has been 
measured at HERA~\cite{Breitweg:2000su} and compared to theoretical 
predictions~\cite{Gordon:1995km,Krawczyk,Fontannaz:2001ek} recently. 
The photoproduction of an isolated prompt photon and a jet has been studied
less extensively~\cite{Breitweg:1997pa}, 
and only partial next-to-leading order codes 
were available up to now~\cite{Gordon:1998yt,Krawczyk}.
However, the presence of 
the jet allows the definition of observables which may provide a more 
detailed picture of the underlying partonic processes than in 
the inclusive case. In particular, it allows to extract information on the 
transverse momentum $k_T$ of the initial state partons. 

The introduction of an effective initial state transverse momentum 
$\langle k_T\rangle$~\cite{Fontannaz:1978uw} 
has become a common practice in 
recent years~\cite{Huston:1995vb,Apanasevich:1998hm,Apanasevich:1999ki} 
and has served 
to bridge large gaps between certain data and NLO theory in fixed target
and hadronic collision experiments of prompt photon production. 
It has been argued that, in addition to the "intrinsic" transverse 
momentum of the initial state partons of a few hundred MeV 
due to the finite size of the proton, initial state soft gluon radiation 
can generate sizeable transverse components of the parton momenta. 
Considerable progress has been made in recent times to account for 
multiple soft gluon emission effects in resummation 
formalisms~\cite{Lai:1998xq}--\cite{werner}, 
but the phenomenological application 
of such calculations has not reached a mature state yet. 
Therefore, simplified phenomenological models for $ k_T$ effects 
have been employed which introduce a Gaussian smearing of the parton 
transverse momenta, the mean value of this effective $k_T$ being 
fitted from data. However, these procedures are rather ad hoc, 
and various experiments lead to
quite different conclusions~\cite{Aurenche:1999gv,Aurenche:2000nz} 
on the necessity to introduce an effective $k_T$. 

\medskip

Recently, the ZEUS collaboration reported on the analysis of prompt photon 
plus jet photoproduction data~\cite{Chekanov:2001aq}. 
From the comparison of the data to results obtained with the Monte Carlo 
program PYTHIA~\cite{pythia}, it has been concluded in~\cite{Chekanov:2001aq}
that a total effective $\langle k_T\rangle$ value of about 1.7\,GeV 
is necessary to fit the data. This value is composed of the mean value of the
intrinsic parton momentum in the proton, $\langle k_T^{\rm{intr}}\rangle$, 
and the parton shower contribution to  
$\langle k_T\rangle$.  The two contributions are combined assuming that
the overall distribution is Gaussian. 
The parton shower contribution was found to be approximately 
1.4\,GeV, and for $\langle k_T^{\rm{intr}}\rangle$ a value of 
$\langle k_T^{\rm{intr}}\rangle\approx 1.25$\,GeV has been fitted.
However, it was found that the same data are already 
well described by HERWIG~\cite{herwig} with a default parton 
intrinsic transverse momentum of zero. 

\medskip 

In this paper, we compare NLO QCD results to the data presented 
in~\cite{Chekanov:2001aq}
and discuss in detail possible sources of differences between theory 
and data. 
In particular, we address the question whether an extra 
effective transverse momentum $\langle k_T\rangle$ -- in addition to the 
one already contained in fixed order NLO QCD 
-- is necessary to fit the ZEUS data. 
We found that our NLO QCD prediction for the quantities which served to
determine $\langle k_T\rangle$ in~\cite{Chekanov:2001aq} describes 
the data without taking into account any extra $k_T$ effects. 
We recall that the
same is true for a large set of prompt photon data in hadronic collisions and
fixed target experiments~\cite{Aurenche:1999gv,Aurenche:2000nz,Binoth:2000qq}. 

Our results have been obtained with a program of partonic event generator 
type which includes the full next-to-leading order corrections to all the 
partonic subprocesses, as well as the box contribution $\gamma g\to\gamma g$.
The program already has been used to study 
inclusive prompt photon photoproduction~\cite{Fontannaz:2001ek}, 
and has been extended 
to allow also the calculation of prompt photon plus jet photoproduction
in the present work.

The paper is organized as follows. 
In section 2, we first describe the main lines of the calculation
our code is based on. Then we discuss the effect of certain kinematic cuts 
and infrared sensitive limits which will be relevant for the comparison 
to the ZEUS data. In section 3 we give numerical results for 
the isolated photon plus jet cross section 
and compare to ZEUS data.  Section 4 contains the study of 
$k_T$-sensitive observables. We show our results for the $p_{\perp}$ 
and $\Delta\phi$ distributions used by ZEUS to determine $\langle k_T\rangle$ 
and discuss their implications.
An appendix contains details of the discussion of symmetric cuts contained 
in section 2.

\section{Description of the method}

\subsection{General setting}

As the general framework of the calculation already has been described in
detail in~\cite{Fontannaz:2001ek}, we will give only a brief overview on the
method here. 

In photoproduction events, the electron acts like a source of 
quasi-real
photons whose spectrum can be described by the Weizs\"acker-Williams 
formula  
\begin{equation}
f^e_{\gamma}(y) = \frac{\alpha_{em}}{2\pi}\left\{\frac{1+(1-y)^2}{y}\,
\ln{\frac{Q^2_{\rm max}(1-y)}{m_e^2y^2}}-\frac{2(1-y)}{y}\right\}\;.
\label{ww}
\end{equation}
The quasi-real photon then either takes part {\em directly} in the 
hard scattering process, or it acts as a composite object, being a 
source of partons which take part in the hard subprocess. 
The latter mechanism is referred to as {\em resolved} process and 
is parametrized by the photon structure functions 
$F_{a/\gamma}(x_{\gamma},Q^2)$. Thus the distribution of partons 
in the electron is a convolution
\begin{equation}
F_{a/e}(x_e,M)=\int_0^1 dy \,dx_{\gamma}\,f^e_{\gamma}(y) \,
F_{a/\gamma}(x_{\gamma},M)\,\delta(x_{\gamma}y-x_e)
\label{resolved}
\end{equation}
where in the "direct" case
$F_{a/\gamma}(x_{\gamma},M)=\delta_{a\gamma}\delta(1-x_{\gamma})$. 

\begin{figure}
\begin{center}
\begin{picture}(100,200)(-15,-130)
\ArrowLine(-30,20)(0,40)
\ArrowLine(-60,20)(-30,20)
\Photon(0,0)(-30,20){2}{5}
\Line(0,0)(30,20)
\Line(0,0)(0,-40)
\Line(0,-40)(-30,-60)
\Photon(0,-40)(30,-60){2}{5}
\BCirc(-32,-62){7}
\Line(-60,-60)(-39,-60)
\Line(-60,-64)(-39,-64)
\Line(-25,-62)(-8,-66)
\Line(-25,-64)(-10,-70)
\Line(-25,-66)(-12,-74)
\Text(0,-30)[t]{direct direct}
%
\ArrowLine(170,20)(200,40)
\ArrowLine(140,20)(170,20)
\Photon(200,0)(170,20){2}{5}
\Line(200,0)(230,20)
\Line(200,0)(200,-40)
\Line(200,-40)(170,-60)
\Gluon(200,-40)(230,-60){2}{5}
\BCirc(234,-63){4}
\Line(238,-60)(260,-62)
\Line(238,-62)(260,-67)
\Photon(238,-65)(260,-79){2}{3}
\BCirc(168,-62){7}
\Line(140,-60)(161,-60)
\Line(140,-64)(161,-64)
\Line(175,-62)(192,-66)
\Line(175,-64)(190,-70)
\Line(175,-66)(188,-74)
\Text(75,-30)[t]{direct fragmentation}
%
%
\ArrowLine(-60,-155)(-30,-135)
\ArrowLine(-90,-155)(-60,-155)
\Photon(-60,-155)(-36,-174){2}{3}
\BCirc(-34,-177){4}
\Line(-30,-178)(-13,-179)
\Line(-30,-176)(-13,-175)
\Gluon(0,-200)(-30,-180){2}{5}
\Line(0,-200)(30,-180)
\Line(0,-200)(0,-240)
\Line(0,-240)(-30,-260)
\Photon(0,-240)(30,-260){2}{5}
\BCirc(-32,-262){7}
\Line(-60,-260)(-39,-260)
\Line(-60,-264)(-39,-264)
\Line(-25,-262)(-8,-266)
\Line(-25,-264)(-10,-270)
\Line(-25,-266)(-12,-274)
\Text(0,-102)[t]{resolved direct}
%
\ArrowLine(140,-155)(170,-135)
\ArrowLine(110,-155)(140,-155)
\Photon(140,-155)(164,-174){2}{3}
\BCirc(166,-177){4}
\Line(170,-178)(187,-179)
\Line(170,-176)(187,-175)
\Gluon(200,-200)(170,-180){2}{5}
\Line(200,-200)(230,-180)
\Line(200,-200)(200,-240)
\Line(200,-240)(170,-260)
\Gluon(200,-240)(230,-260){2}{5}
\BCirc(234,-263){4}
\Line(238,-260)(260,-262)
\Line(238,-262)(260,-267)
\Photon(238,-265)(260,-279){2}{3}
\BCirc(168,-262){7}
\Line(140,-260)(161,-260)
\Line(140,-264)(161,-264)
\Line(175,-262)(192,-266)
\Line(175,-264)(190,-270)
\Line(175,-266)(188,-274)
\Text(75,-102)[t]{resolved fragmentation}
\end{picture}
\end{center}
\caption{Examples of contributing subprocesses at leading order}
\label{fig1}
\end{figure}
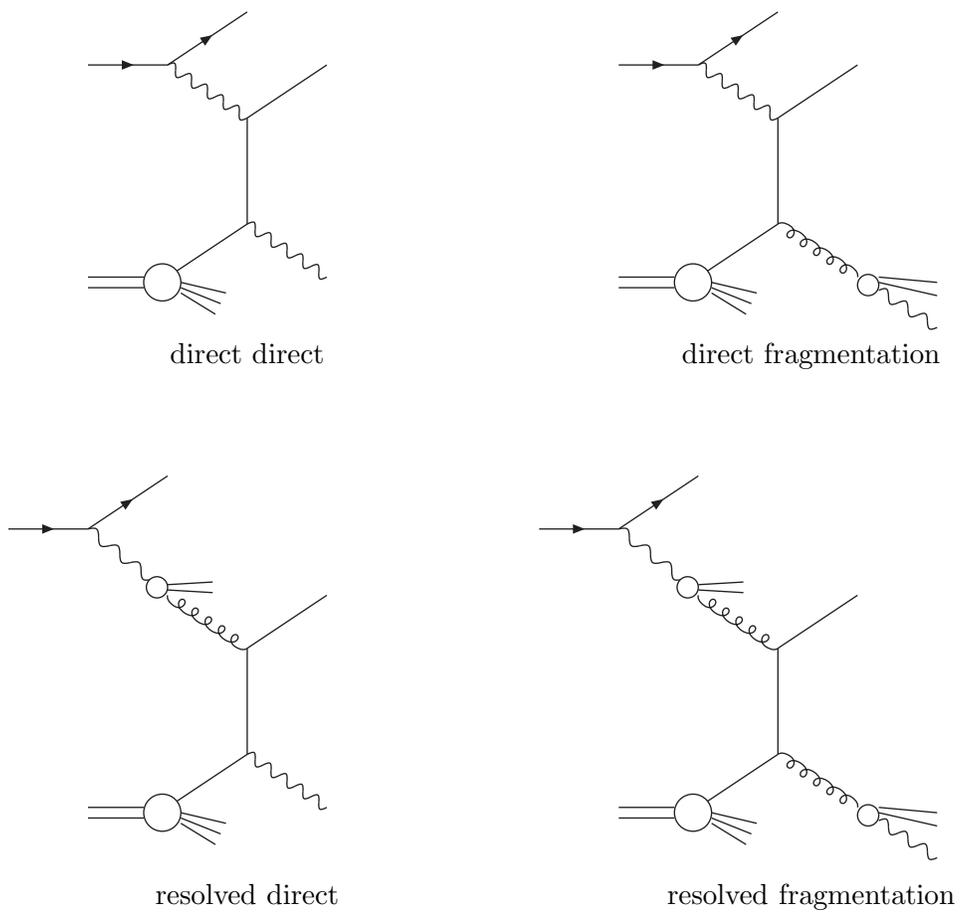

Similarly, a high-$p_T$ photon in the final state can either originate 
directly from the  hard scattering process or it can be produced by the
fragmentation of a hard parton emerging from the hard scattering process. 
Thus we distinguish four categories of production mechanisms, as illustrated in 
fig.\,\ref{fig1}: 
1.\,direct direct \, 2.\,direct fragmentation 
\, 3.\,resolved direct \, 4.\,resolved fragmentation.  
We implemented the full set of next-to-leading order corrections 
to all four subprocesses. 
Note that the photon structure functions behave for large $Q^2$ 
as 
$F_{a/\gamma}(x,Q^2)\sim \ln{Q^2/\Lambda^2}\sim 1/\alpha_s(Q^2)$. 
An analogous argument holds for the fragmentation functions, 
such that a consistent NLO calculation requires the inclusion of the 
full ${\cal O}(\alpha_s)$ corrections not only to the 
"direct direct" part but also to the resolved and/or fragmentation
parts\footnote{In the following, we will denote by "resolved" the sum of
resolved direct and resolved fragmentation parts and by "direct" the sum of 
direct direct and direct fragmentation parts.}. 
The matrix elements for these 
NLO corrections have been taken from the 
literature~\cite{Aurenche:1984hc,ellissexton,Aurenche:1987ff}.
We also included the box contribution 
$g\gamma \to g\gamma$, which is formally NNLO, 
but known to be sizeable~\cite{Fontannaz:1982et,Aurenche:1992sb}.

It is well known that at NLO the attribution of certain finite terms 
to either of the four categories listed above 
is scheme dependent and therefore only the sum 
of all parts has a physical meaning. 

In order to isolate the infrared singularities appearing in the calculation 
at next-to-leading order, a modified phase-space slicing method has been 
used~\cite{Fontannaz:2001ek,Binoth:2000qq}.  

For the definition of jets, the $k_T$--algorithm~\cite{ktalgo} has been 
employed, 
as has been done by the ZEUS collaboration. 

To single out the prompt photon events from the background
of secondary photons produced by the decays of light mesons,
isolation cuts have to be imposed on the photon signals in the experiment.
We use the following isolation criterion:
In a cone $C$ centered around the photon direction 
in the rapidity and azimuthal angle plane, defined by 
$$C=\left\{(\eta,\phi)|\left( \eta -\eta_{\gamma} \right)^{2}+
\left(\phi-\phi_{\gamma} \right)^{2} \leq  R^{2}\right\}\;,$$ 
the amount of deposited hadronic transverse energy
$E_T^{had}(R)$ is required to be smaller than 
a fraction $\epsilon_c$ of the photon transverse momentum. 
We use $\epsilon_c=0.1$ and $R$ = 1 to match the cuts of the ZEUS
collaboration.

Since isolation cuts impose additional phase space restrictions, the question
has been raised if factorization properties valid for the inclusive case 
might be spoiled by isolation. This issue is discussed in detail 
in~\cite{Catani:1998yh,CFG}. 
Here we just note that the above criterion, being based on 
boost-invariant {\em transverse} energies, leads to infrared safe cross
sections.

\subsection{Effects of kinematic cuts, infrared sensitive limits}\label{cut}

As a consequence of phase space restrictions induced 
for example by kinematic cuts, 
certain observables calculated at fixed order 
may show instabilities (integrable singularities) at some critical point  
of phase space~\cite{Catani:1997xc,Catani:1998yh}. 
Examples  will be discussed below.  

\subsubsection*{Symmetric cuts on $E_{T}^{jet}, E_{T}^{\gamma}$}

\begin{figure}[htb]
\begin{center}
\mbox{\epsfig{file=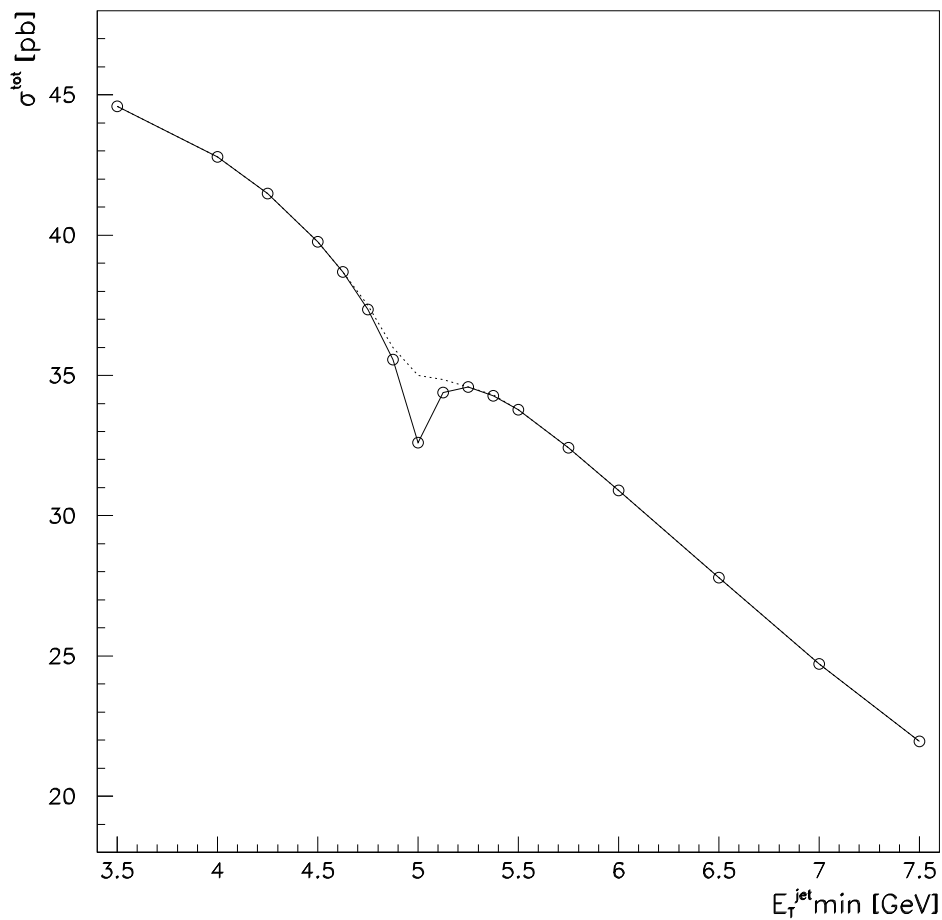,height=14cm}}
\end{center}
\caption{Nonisolated total $\gamma+$\,jet cross section for 
$E_T^{\gamma}> 5$\,GeV  as a function of 
$E_{T,\rm{min}}^{jet}$. The photon energy range is $0.2<y<0.9$, 
the rapidities are integrated over 
$-1.5<\eta^{jet}<1.8$ and $-0.7<\eta^{\gamma}<0.9$.} 
\label{etcut}
\end{figure}

In order to restrict both prompt photons and jets to well-measured 
kinematic regions, the ZEUS collaboration has required 
$E_T^{\gamma}> 5$\,GeV and $E_T^{jet}> 5$\,GeV in their 
analysis~\cite{Chekanov:2001aq}. However, 
as it is well known~\cite{Aurenche:1997im,Frixione:1997ks}, 
symmetric cuts on the transverse energies should be avoided 
since they select a region where fixed order NLO QCD loses its
predictive power. 
This can be seen by considering the total cross section as a 
function of $E_{T,\rm{min}}^{jet}$ as shown in  
fig.\,\ref{etcut}. The cut on the transverse energy of the photon 
has been fixed to 
$E_{T,\rm{min}}^{\gamma}=5$\,GeV. The figure shows that the  
 NLO QCD prediction varies rapidly around the 
point $E_{T,\rm{min}}^{jet}=E_{T,\rm{min}}^{\gamma}$. 
In particular, it is striking that the cross section 
is {\em not} immediately monotonically decreasing for 
increasing $E_{T,\rm{min}}^{jet}$.
The reason for the dip at $E_{T,\rm{min}}^{jet}=E_{T,\rm{min}}^{\gamma}$
is that at this point the phase space for the unobserved parton in the real
corrections is severely restricted, 
such that the (negative) virtual  corrections are not sufficiently 
balanced by the (positive) real corrections\footnote{We 
checked by varying the phase space slicing parameter that this effect is 
not an artifact of the slicing method.}. 
As a result we observe a violation of unitarity in the range 
5.0\,GeV$\leq E_{T,\rm{min}}^{jet}\leq 5.25$\,GeV, where the cross section 
increases when the available phase space decreases. 
This phenomenon, that we discuss in detail in the Appendix, 
is an artifact of the fixed order calculation. In this particular 
region around $E_{T,\rm{min}}^{jet}=E_{T,\rm{min}}^{\gamma}$, the 
higher order corrections should be resummed to produce 
a curve that should look like the dashed curve in fig.\,\ref{etcut}. 
This resummation is not considered in this paper 
(but the large logarithms responsible for this behaviour are 
identified, see  Appendix A) because the unphysical effect at 
$E_{T,\rm{min}}^{jet}=E_{T,\rm{min}}^{\gamma}$ is in general numerically weak. 
However, we shall explore the effect of different choices of 
$E_{T,\rm{min}}^{jet}$ when calculating the cross section 
corresponding to ZEUS data. It is clear that an asymmetric cut
in the experiment would avoid any trouble. 

\clearpage

\subsubsection*{Definition of $x_{\gamma}$, limit $x_{\gamma}\to 1$}

\begin{figure}[htb]
\begin{center}
\mbox{\epsfig{file=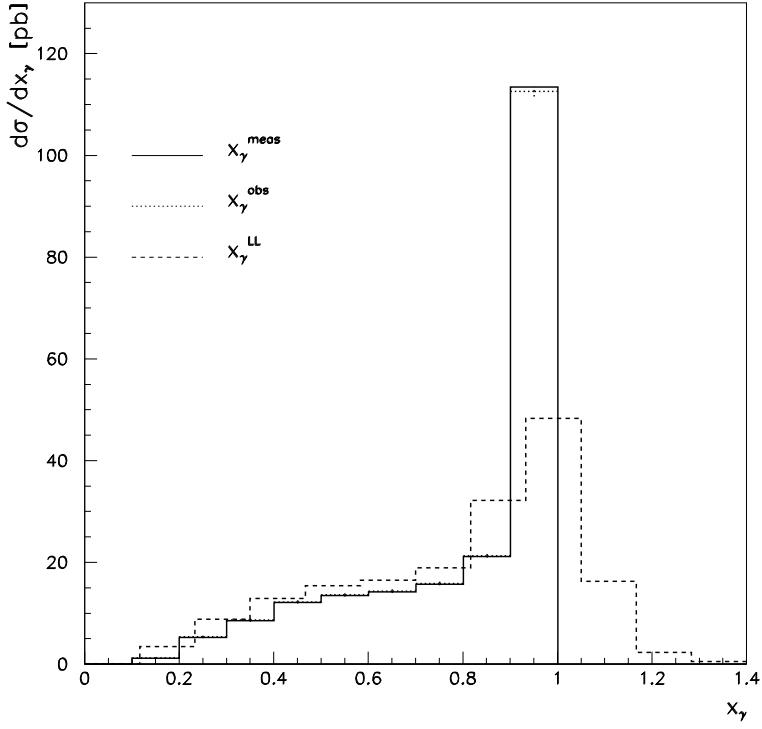,height=11cm}}
\end{center}
\caption{Behaviour of $d\sigma/dx_{\gamma}^{LL}$, 
$d\sigma/dx_{\gamma}^{meas}$ and $d\sigma/dx_{\gamma}^{obs}$ 
for the isolated $\gamma$ + jet cross section.}
\label{xll}
\end{figure}

Care also has to be taken in the treatment of the variable $x_{\gamma}$
near the  boundary $x_{\gamma}\to 1$. 
At leading order, the fraction $x_{\gamma}$ of the incoming photon longitudinal 
momentum is defined as 
\begin{equation}
x_{\gamma}=\frac{E_{T3}(e^{-\eta_{3}}+e^{-\eta_{4}})}{2E_{\gamma}}
\end{equation}
where $\eta_{3}$ and $\eta_{4}$ are the rapidities of the two outgoing 
partons and $E_{\gamma}=yE_e$. 
At next-to-leading order, the "true" $x_{\gamma}$ for 
2 $\to$ 3 processes is given by 
$x_{\gamma}=\sum_{i=3}^5(E_{Ti}e^{-\eta_{i}})/(2E_{\gamma})$, 
but since particle 5 is not observed, $x_{\gamma}$ is no 
longer an experimentally accessible observable. 
Hence, to obtain an estimate of the longitudinal momentum fraction of 
the incoming photon, the variable $x_{\gamma}^{obs}$, defined as
\begin{equation}
x_{\gamma}^{obs}=\frac{E_T^{\gamma}e^{-\eta^{\gamma}}+
E_T^{jet}e^{-\eta^{jet}}}{2E_{\gamma}}\label{xobsth}
\end{equation} 
is frequently used.
However, the above definition of $x_{\gamma}^{obs}$ may lead to infrared
sensitive cross sections~\cite{Aurenche:2000nc}. Fixing $E_T^{\gamma}$ 
{\em and} $E_T^{jet}$ strongly constrains the phase space of parton 5 
when $x_{\gamma}^{obs}$ goes to one. This is reflected by 
$\log(1-x_{\gamma}^{obs})$ terms generated by the higher order corrections. 
In general, $x_{\gamma}^{obs}$ is integrated over, typically in the range 
$0.75\le x_{\gamma}^{obs}\le 1$, such that the singular behaviour is 
sufficiently smoothed out for a fixed order calculation to remain valid. 
However, if the width of the last bin at 
$x_{\gamma}^{obs}\to 1$ is small, the shape of the 
theoretical prediction near $x_{\gamma}^{obs}= 1$ 
is considerably dependent on the bin width. 
For this reason, the variable 
$x_{\gamma}^{LL}=E_T^{\gamma}(e^{-\eta^{\gamma}}+
e^{-\eta^{jet}})/(2E_{\gamma})$,  
which has a smoother behavior for $x_{\gamma}\to 1$, 
has been proposed in~\cite{Aurenche:2000nc}. 
It also has the advantage that it
does not depend on  $E_T^{jet}$ and thus does not contain an uncertainty
from the jet energy reconstruction. Note that $x_{\gamma}^{LL}$ can be bigger 
than one, as illustrated in fig.\,\ref{xll}. 

For the analysis presented in~\cite{Chekanov:2001aq}, the ZEUS collaboration 
did not use $x_{\gamma}^{obs}$, but the variable $x_{\gamma}^{meas}$,
defined as
\begin{equation}
x_{\gamma}^{meas}=\frac{\sum_{\gamma,i}(E-p_z)}{2E_{e}\,y^{JB}}
=\frac{E_T^{\gamma}e^{-\eta^{\gamma}}+
\sum_{i}E_{Ti}e^{-\eta^{i}}}{2E_{e}\,y^{JB}}
\label{xmeas}
\end{equation}
where the sum is over the prompt photon and the contents of the jet, 
and $y^{JB}$ is the energy fraction of the incoming photon, 
reconstructed  with the Jacquet-Blondel method: 
\begin{equation}
y^{JB}=\frac{\sum(E-p_z)}{2E_e}
\label{yjb}
\end{equation}
Here the sum is over all energy-flow objects in the event, each of which is
treated as due to a massless particle with energy $E$ and 
longitudinal momentum component $p_z$.
From an experimental point of view, $x_{\gamma}^{meas}$ is convenient
since a lot of calorimeter calibration
systematics cancel out. The difference between definitions  
(\ref{xobsth}) and (\ref{xmeas}) for  $x_{\gamma}$ depends on the 
definitions of $E_T^{jet}$ and $\eta^{jet}$ in (\ref{xobsth}). 
If the Snowmass conventions 
\begin{equation}
E_T^{jet}=\sum_i E_{Ti}\;,\;
\eta^{jet}=\frac{\sum_i E_{Ti}\eta^i}{E_T^{jet}}\;,\;
\phi^{jet}=\frac{\sum_i E_{Ti}\phi^i}{E_T^{jet}}
\label{fisnow}
\end{equation}
are used, one has 
$$x_{\gamma}^{meas}-x_{\gamma}^{obs}=\frac{1}{2E_{\gamma}}\left\{
\sum_i E_{Ti}e^{-\eta^{i}}-
\left(\sum_i E_{Ti}\right)e^{-\frac{\sum_i E_{Ti}\eta_i}{E_T^{jet}}}\right\} 
\ge 0\;.$$
For our analysis, where maximally two massless partons form a jet,
the difference between $x_{\gamma}^{obs}$ and $x_{\gamma}^{meas}$ is
numerically indistinguishable (see fig.\,\ref{xll}).

Obviously, what has been said above on the limit $x_{\gamma}^{obs}\to 1$
is valid for $x_{\gamma}^{meas}\to 1$ as well.

\clearpage

\section{Numerical results and comparison to ZEUS data} 
 
For the numerical studies we use the 
rapidity cuts $-1.5<\eta^{jet}<1.8$ and $-0.7<\eta^{\gamma}<0.9$  
chosen by ZEUS to restrict the photon and the jet to well measured kinematic
regions.
We also use the cuts $E_{T,\rm min}^{jet}=E_{T,\rm min}^{\gamma}=5$\,GeV
in order to match the ZEUS cuts, although such symmetric cuts 
select a region where fixed order perturbative QCD shows 
instabilities, as discussed 
in detail in section \ref{cut} and in the appendix. 
From fig.~\ref{etcut} one can estimate that the uncertainty 
in the theoretical prediction introduced by using these cuts is 
of the order of 6-10\% for the total cross section.

We set $Q^2_{\rm max}=1$\,GeV$^2$ for the virtuality of the photons in
(\ref{ww}) and restrict the photon energy to the range $0.2<y<0.7$ as 
reconstructed by ZEUS.  
We take the MRST2~\cite{Martin:2000ww} 
parametrization for the parton distributions in the proton,  
AFG~\cite{Aurenche:1994in} for the parton distributions in the photon, 
BFG~\cite{Bourhis:1998yu} for the fragmentation functions into photons,  
$n_f=4$ flavors, and for $\alpha_s(\mu)$ we use an exact 
solution of the two-loop renormalization group
equation, and not an expansion in log$(\mu/\Lambda)$. 
Unless stated otherwise, the scale choice $\mu=p_T^{\gamma}$ 
has been adopted and $M$ and $M_F$ have been set equal to $\mu$. 
The rapidities refer to the $e\,p$ laboratory frame, with the HERA
convention that the proton is moving towards positive rapidity. 

\begin{figure}[htb]
\begin{center}
\mbox{\epsfig{file=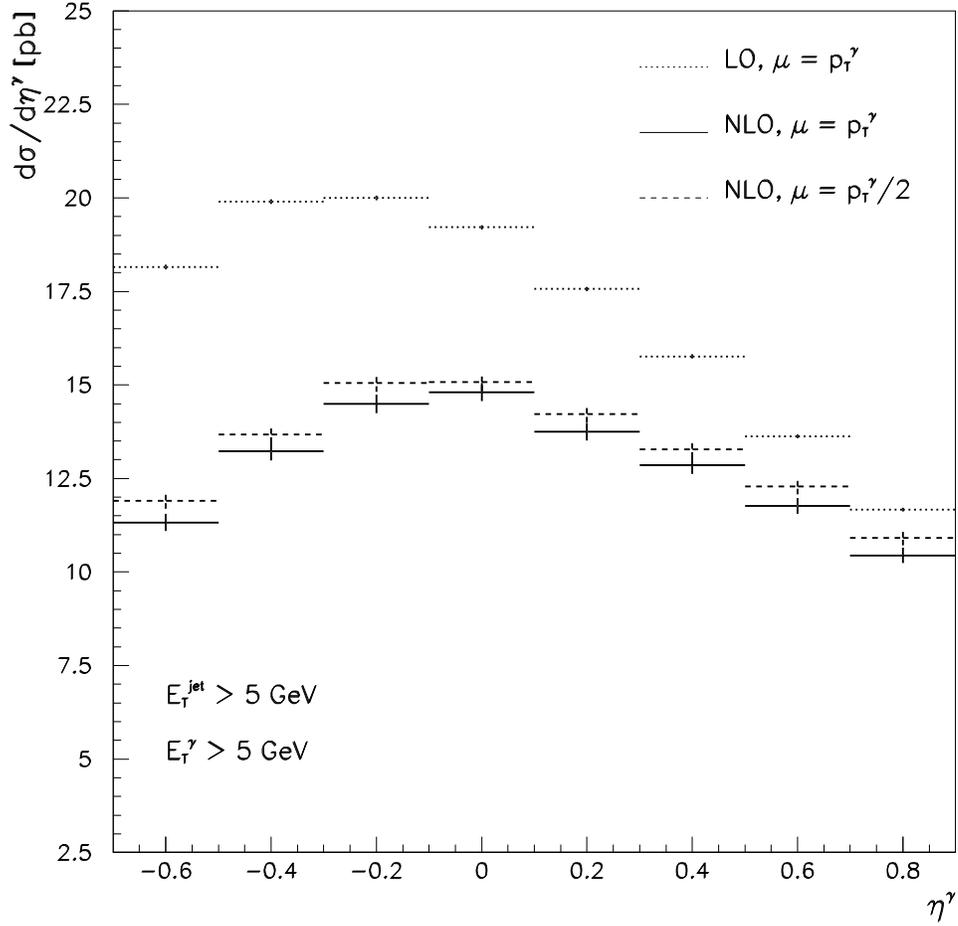,height=14cm}}
\end{center}
\caption{Magnitude of leading order and NLO results for the isolated 
cross section $d\sigma^{\gamma+jet}/d\eta^{\gamma}$. Isolation with 
$\epsilon_c=0.1, R=1$, jet rapidities integrated over $-1.5<\eta^{jet}<1.8$\,.}
\label{bornnlo}
\end{figure}

\begin{figure}[htb]
\begin{center}
\mbox{\epsfig{file=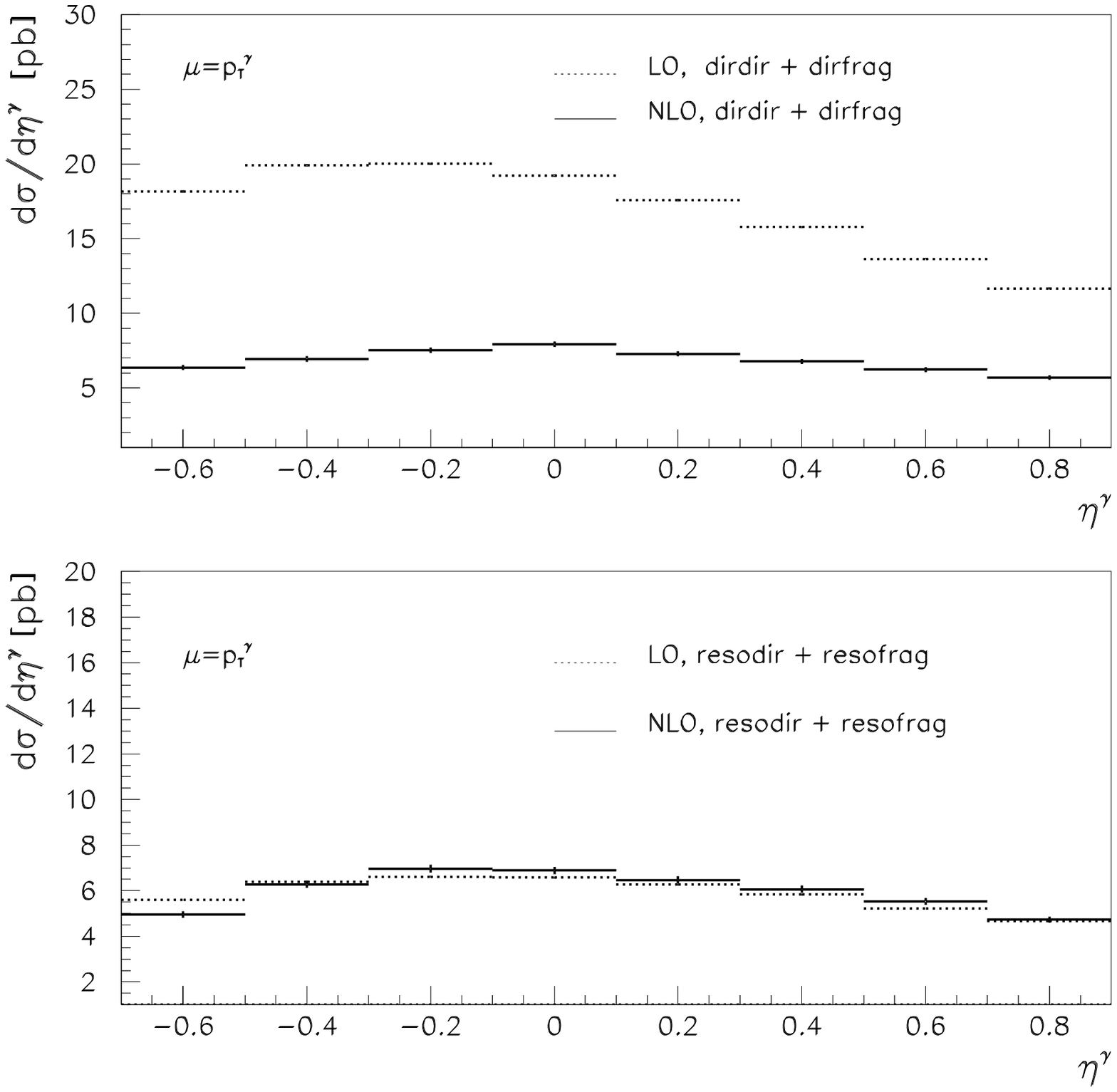,height=14cm}}
\end{center}
\caption{Magnitude of LO and NLO contributions for resolved and direct 
parts  
separately, with isolation ($\epsilon_c=0.1, R=1)$.}
\label{bornnlosep}
\end{figure}

Fig.\,\ref{bornnlo} shows the relative magnitude of the NLO 
result to the leading order\footnote{By "leading order" we mean the sum of the
lowest order terms of all four subprocesses direct direct, 
direct fragmentation, resolved direct and resolved fragmentation.} 
result  for the isolated photon rapidity distribution. 
As the box contribution has the same kinematics as the leading order 
direct direct part, we included the box contribution in this part, 
such that "direct direct" always means direct direct plus box.
Since the higher order corrections 
to the direct direct term are negative and those to the resolved 
part are small (for the scale choice $\mu=p_T^{\gamma}$, 
see fig.\,\ref{bornnlosep}), the overall NLO result is 
lower than the leading order result. 
We note that the higher order corrections 
to the direct  part are most negative for 
$E_{T,\rm{min}}^{jet}= E_{T,\rm{min}}^{\gamma}$, for asymmetric cuts the 
absolute value of the negative contributions decreases.

\begin{figure}[htb]
\begin{center}
\mbox{\epsfig{file=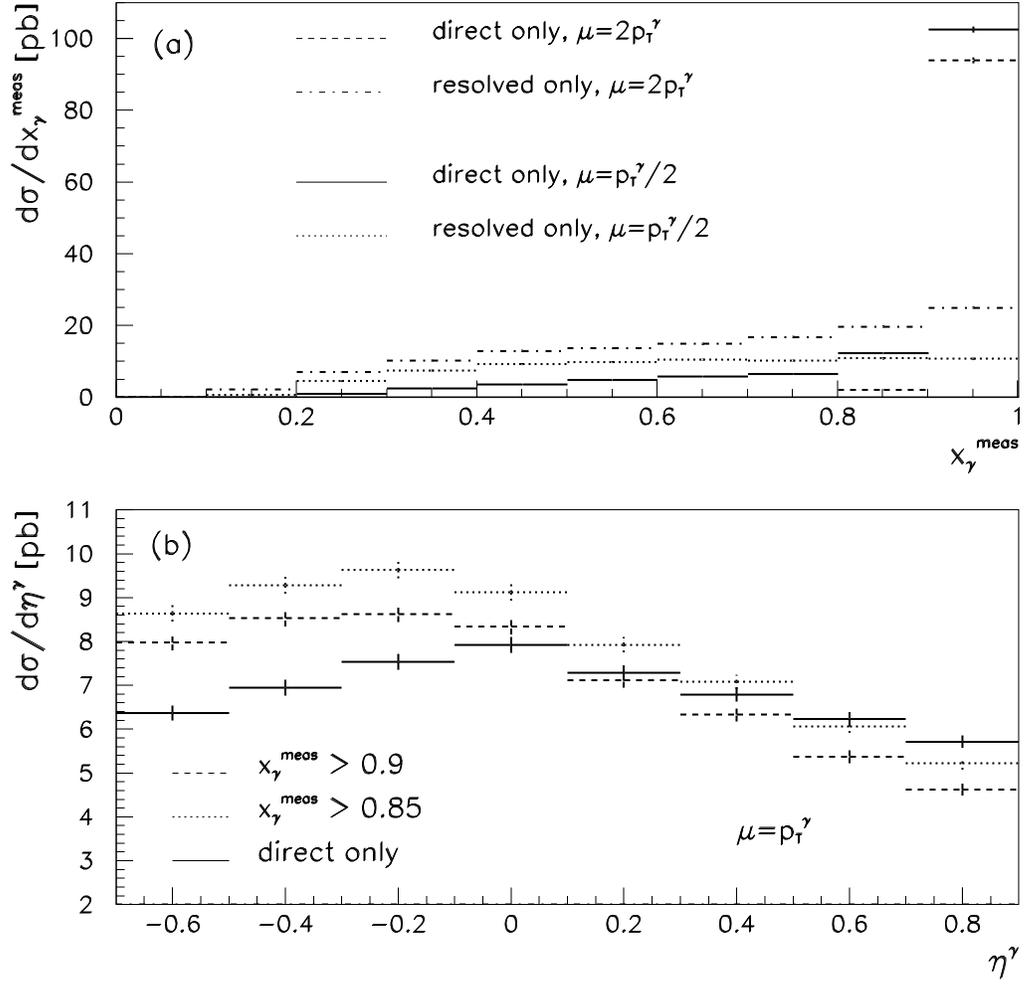,height=15cm}}
\end{center}
\caption{(a) $d\sigma/dx_{\gamma}^{meas}$ for direct and resolved components 
separately and two different scale choices, 
(b) $d\sigma/d\eta^{\gamma}$ for direct component (no cut) compared 
to the full result with the cut 
$x_{\gamma}^{meas}>0.9\,(0.85)$.}
\label{dirxcut}
\end{figure}

In order to single out the events stemming from a direct photon in the 
initial state, the cut $x_{\gamma}^{meas}>0.9$ has been applied 
by ZEUS. However, it has to be kept in mind that beyond leading order,
the direct part has a non-negligible tail away from $x_{\gamma}^{meas}=1$, 
and that the contributions from the resolved part to the  bin  
$x_{\gamma}^{meas}>0.9$ are rather large, as can be seen from 
Fig.\,\ref{dirxcut}\,(a). 
Fig.\,\ref{dirxcut}\,(b) compares the  sum  direct + resolved  with 
the cut $x_{\gamma}^{meas}>0.9\,(0.85)$
to  direct direct  
only with no cut on $x_{\gamma}^{meas}$ 
for the photon rapidity distribution. 
One observes that there is a significant difference in the shape 
between the direct part and the result obtained by imposing the cut  
$x_{\gamma}^{meas}>0.9$ to the full set of subprocesses, the latter 
being higher in the backward region and lower in the forward region. 
Hence imposing a stringent cut on $x_{\gamma}^{meas}$ cannot be considered 
as fully equivalent to singling out the direct events only.

\clearpage

\subsection{Comparison of photon rapidity and $x_{\gamma}$ distributions to
ZEUS data}

\begin{figure}[hb]
\begin{center}
\mbox{\epsfig{file=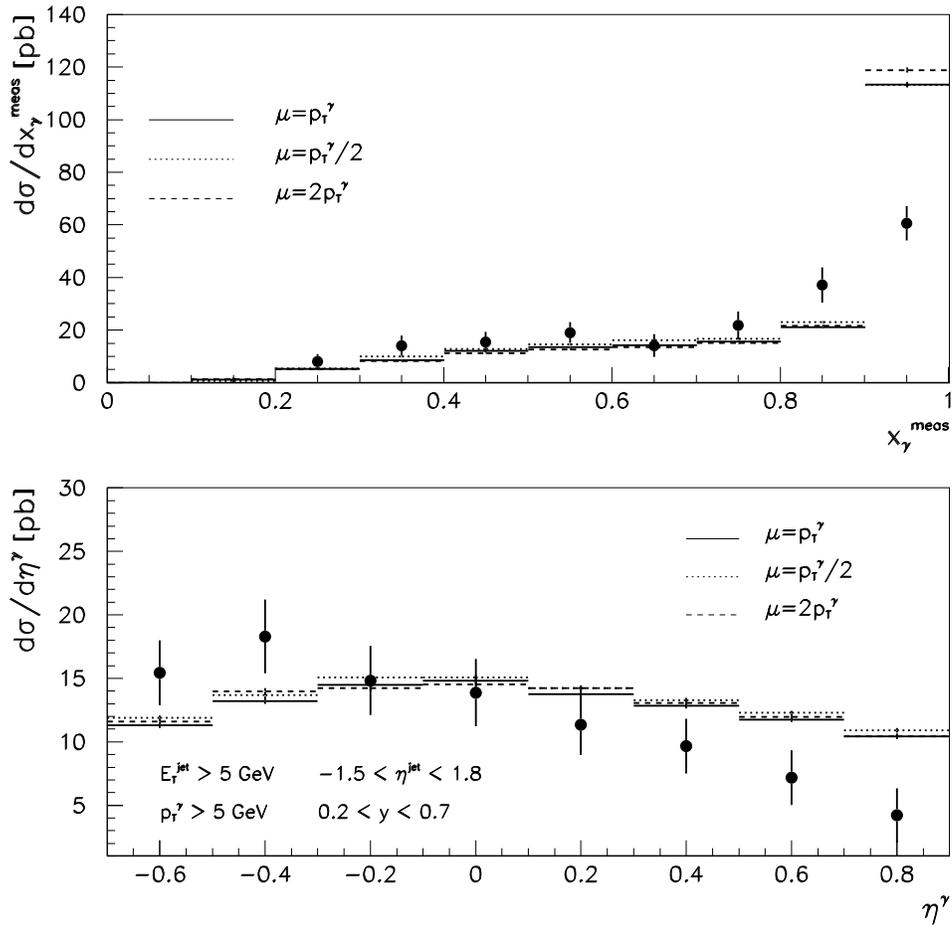,height=14cm}}
\end{center}
\caption{$d\sigma^{\gamma+jet}/dx_{\gamma}^{meas}$   
and $d\sigma^{\gamma+jet}/d\eta^{\gamma}$ for the 
 scale choices $\mu=p_T^{\gamma}$ (default), 
$\mu=p_T^{\gamma}/2$ and $\mu=2p_T^{\gamma}$  
(always $M_F=M=\mu$) together with ZEUS data.  
The errors on the data are statistical only.}
\label{xgaobs}
\end{figure}

Fig.\,\ref{xgaobs} shows the distributions   
$d\sigma/dx_{\gamma}^{meas}$  and the photon rapidity 
distribution for various scale choices together with the ZEUS data. 
One can see that the theoretical uncertainty due to 
overall scale variations, i.e. 
varying $\mu$ between $p_T^{\gamma}/2$ and $2p_T^{\gamma}$  
and setting $M_F=M=\mu$,  
is very small.  
For the distribution in $x_{\gamma}^{meas}$  one observes 
that theory overshoots the data substantially in the last bin, whereas
the data are described well in the remaining bins. 
In  $d\sigma/d\eta^{\gamma}$ there is a rather large 
discrepancy between theory and data in the forward region.

\begin{figure}[htb]
\begin{center}
\mbox{\epsfig{file=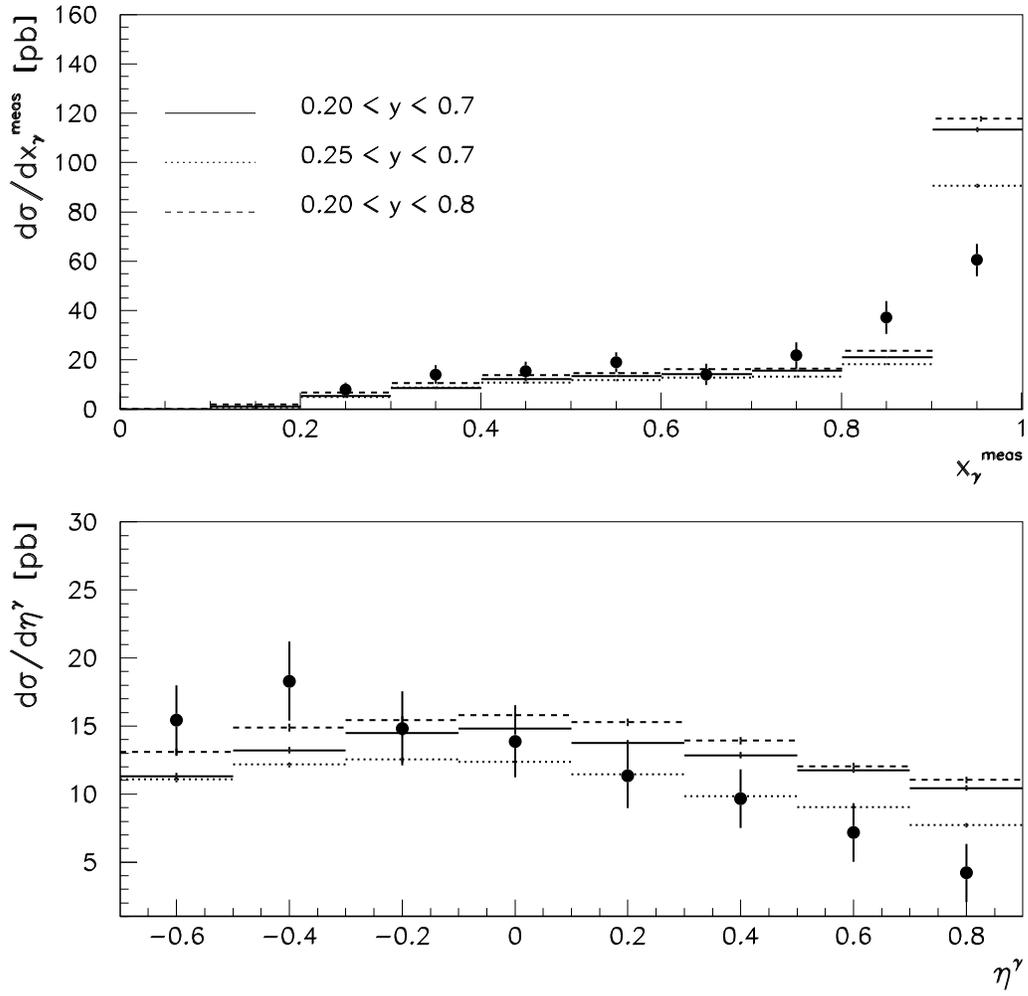,height=15cm}}
\end{center}
\caption{$d\sigma^{\gamma+jet}/dx_{\gamma}^{meas}$ 
and photon rapidity distribution 
$d\sigma^{\gamma+jet}/d\eta^{\gamma}$  for 
 different cuts on the photon energy: 
$0.2<y<0.7$ (default),  $0.25<y<0.7$,  $0.2<y<0.8$.}
\label{ymin}
\end{figure}

Reasons for these discrepancies may be the following:
\begin{enumerate}
\item
The data are at the detector level, i.e. they are not fully corrected 
for energy losses in the detector and other effects. Therefore 
data corrected to the hadron level may differ in a non-negligible way 
from the ones displayed in fig\,\ref{xgaobs}. 
Moreover our calculations are performed at the parton level and hadronization
effects may also lead to sizeable corrections at these low values of
$p_T^{jet}$ (see also point 4. below).
\item There is an uncertainty in the reconstruction of the 
"true" photon energy $y$ with the Jacquet-Blondel method 
(cf. eq.~(\ref{yjb})).  
In general, corrections for detector effects and 
energy calibration tend to shift the range of $y$ to 
slightly higher values as compared to the original range of $y^{JB}$. 
To study the effect of an uncertainty in $y$, 
we  varied the cuts on $y$, increasing the lower cut 
from $y_{min}=0.2$ to $y_{min}=0.25$, respectively increasing the upper cut
from $y_{max}=0.7$ to $y_{max}=0.8$, as shown in fig.\,\ref{ymin}.  
\item  Concerning $d\sigma/dx_{\gamma}^{meas}$: \\  
There is an uncertainty in the energy and $p_z$ values
 entering in the definition (\ref{xmeas})
of $x_{\gamma}^{meas}$. Taking into account this uncertainty 
in the  calculation would introduce a Gaussian smearing in $E$
and $p_z$ which would smooth out the theoretical curve near 
$x_{\gamma}^{meas}\to 1$. 
\item 
As already discussed in~\cite{Fontannaz:2001ek},  
a non-negligible part of the hadronic energy 
measured in the isolation cone may be due to underlying events, 
especially in the forward region. 
Therefore it is possible that experimentally, the isolation cut 
removes events which pass in the theoretical (parton level) 
simulation since there the underlying event contamination is not present. 
It has been estimated~\cite{Fontannaz:2001ek} that this 
can have an effect of the order of 30\% in the forward region. 
\end{enumerate}

\begin{figure}[htb]
\begin{center}
\mbox{\epsfig{file=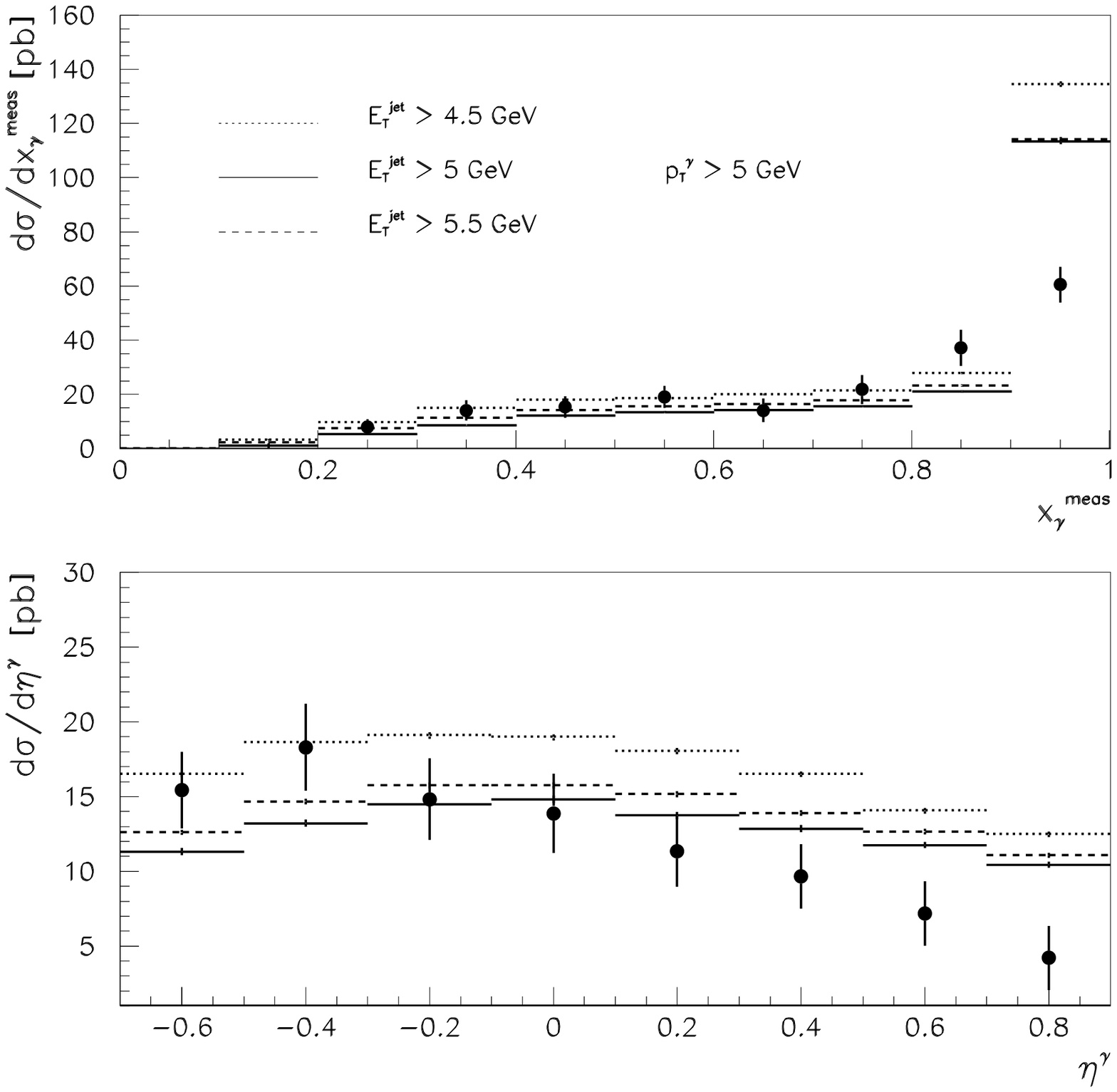,height=15cm}}
\end{center}
\caption{ $d\sigma^{\gamma+jet}/dx_{\gamma}^{meas}$ and 
$d\sigma^{\gamma+jet}/d\eta^{\gamma}$ for different cuts on $E_{T}^{jet}$.}
\label{jetmin}
\end{figure}

Fig.\,\ref{jetmin} again demonstrates (see also fig.\,\ref{etcut})
that the cross section is very sensitive to a 
variation of $E_{T,\rm{min}}^{jet}$ around the value 
$E_{T,\rm{min}}^{jet}=E_{T,\rm{min}}^{\gamma}$. 
It can be clearly seen that increasing the cut on 
$E_T^{jet}$ from 5\,GeV to 5.5\,GeV 
does {\em not} lead to a decrease in the cross section, as explained 
in section \ref{cut}. 
For the rapidity distribution  
the theoretical uncertainty introduced by using symmetric cuts 
can be as large as 30\%. 

\medskip

An intrinsic $\langle k_T\rangle$ would not improve the agreement 
between theory and data for the photon rapidity distribution,
since it would shift the theoretical curve upwards without changing its
shape. We note also that the data 
for $d\sigma/d\eta^{\gamma}$ 
and $d\sigma/dx_{\gamma}^{meas}$ 
are in general well described by PYTHIA 6.129 with default 
$\langle k_T\rangle$ values~\cite{Chekanov:2001aq}. 
Only in the backward region in the photon rapidity distribution, 
PYTHIA is lower than the data, 
the same trend is visible in the NLO theory curve. 

\begin{figure}[htb]
\begin{center}
\mbox{\epsfig{file=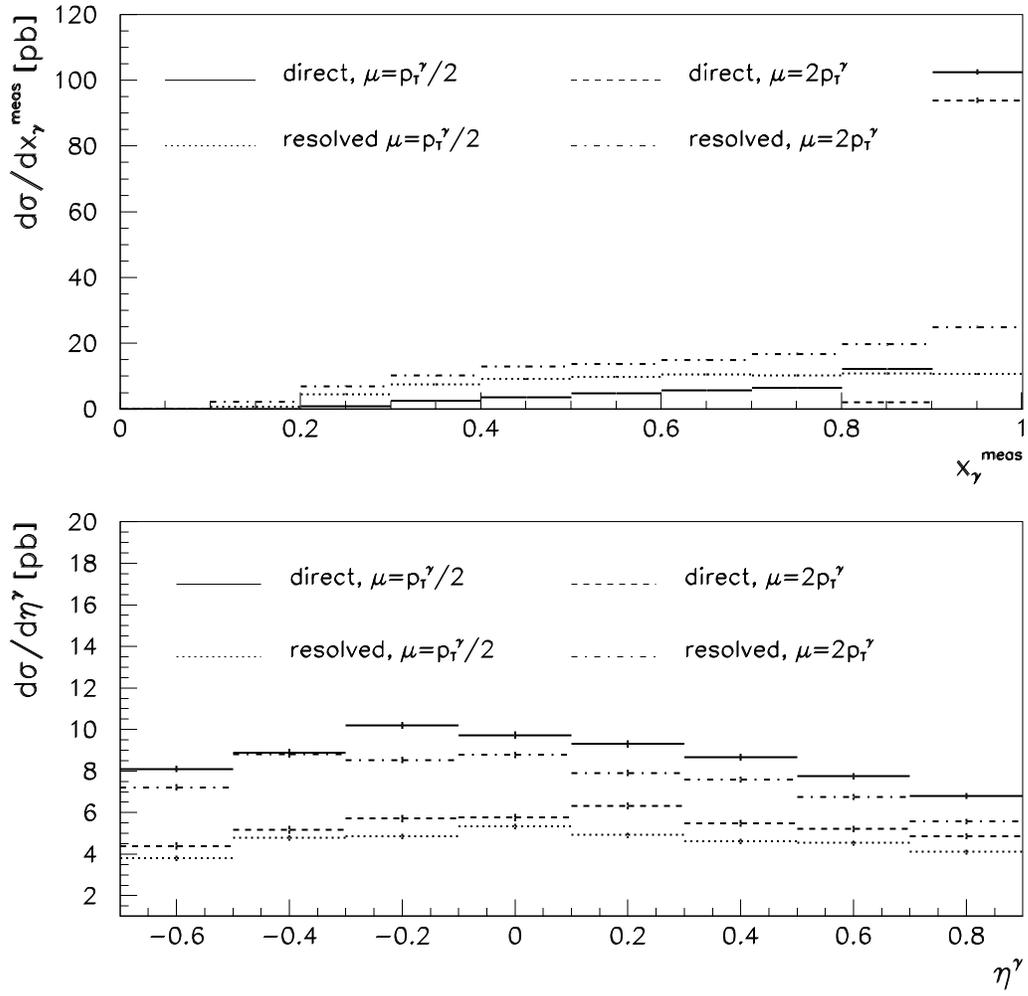,height=15cm}}
\end{center}
\caption{Contributions from resolved and direct initial state photons
separately for two different scale choices.}
\label{scaledep}
\end{figure}

Figure \ref{scaledep} shows the contributions from direct and resolved 
initial state photons separately. 
As the those parts separately are highly scale dependent, 
results for two ("extreme") scale choices are shown. The figure 
for the rapidity distribution illustrates that the statement 
which part -- i.e. resolved or direct part -- 
dominates in the forward region, is scale dependent.

\clearpage


\section{Results for $k_T$ - sensitive observables and
comparison to ZEUS data}

In this section we compare our results 
with the ZEUS data for the $k_T$ - sensitive
observables $p_{\perp}$ and $\Delta\phi$  which served 
to determine $\langle k_T\rangle$ in~\cite{Chekanov:2001aq}. 
The quantity $p_{\perp}$ is the momentum component of the photon 
perpendicular to the jet direction, defined as
\begin{equation}
 p_{\perp}=|\vec{p}_T^{\,\gamma}\times\vec{p}_T^{\,jet}|/p_T^{jet}
 \label{pperp}
\end{equation}
Here $p_T^{jet}$  is {\em not} the Snowmass
$E_T^{jet}$, but the modulus of the transverse components of the 
vector sum of the particle momenta which form the jet. 
$\Delta\phi$ is the azimuthal acollinearity between the photon and the jet.
To define $\Delta\phi$, the Snowmass
definition (see eq.~(\ref{fisnow})) for the jet azimuthal angle has been 
used by ZEUS.
In the experiment, a cut of $x_{\gamma}^{meas}>0.9$ has been imposed in 
order to select  events which are 
predominantly from direct incoming photons, thereby suppressing eventual 
contributions to  $\langle k_T\rangle$ from the resolved photon. 
The data for these quantities have been corrected to hadron level. 
As the value of $\langle k_T\rangle$
affects mainly the shape of the distributions, the cross
sections have been normalized to one in order to 
minimize uncertainties in the calorimeter energy scale.   

\medskip

\begin{figure}[htb]
\begin{center}
\mbox{\epsfig{file=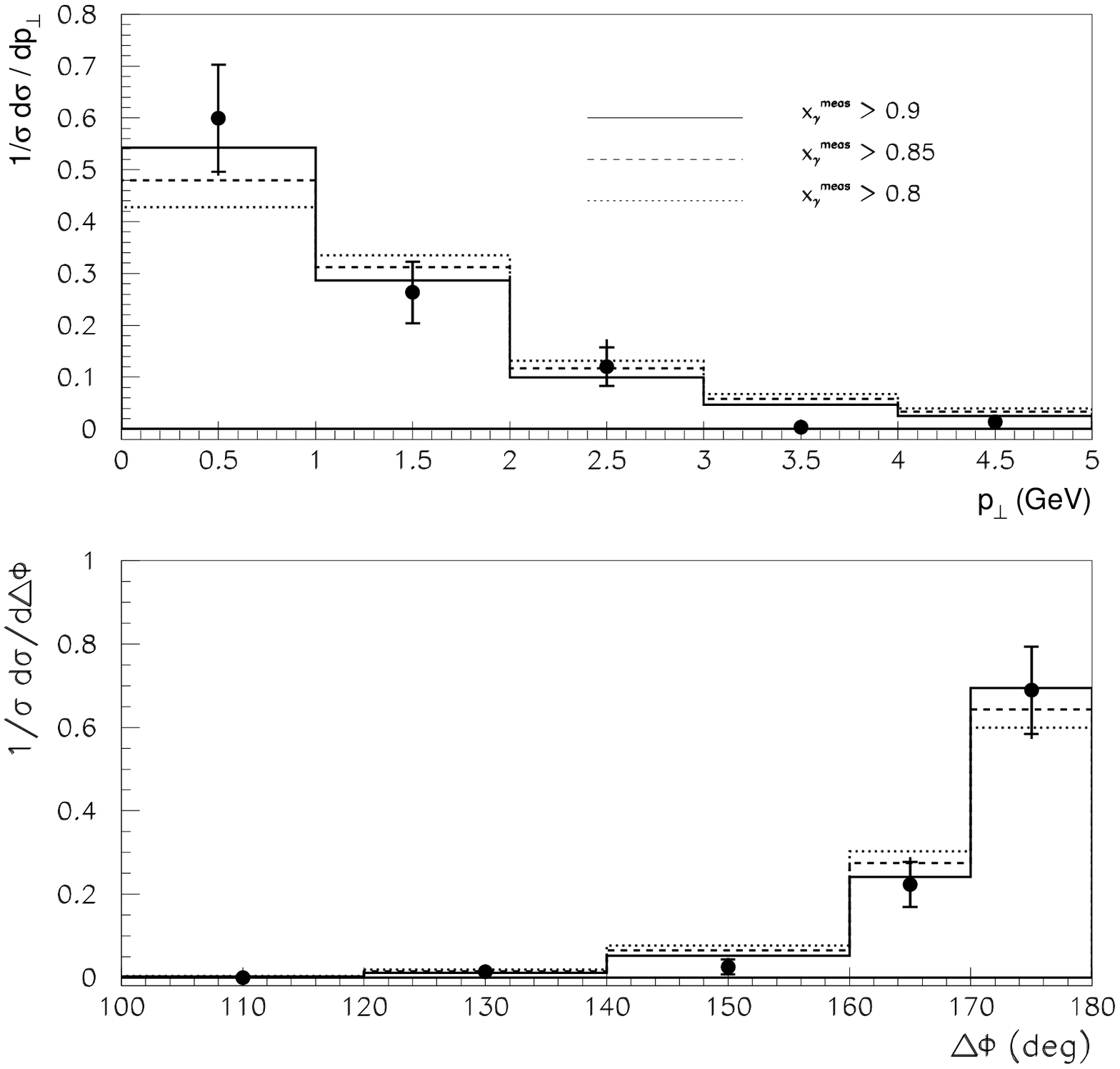,height=15cm}}
\end{center}
\caption{Normalized cross section differential in $p_{\perp}$ and $\Delta\phi$
for  $x_{\gamma}^{meas}> 0.8, 0.85, 0.9$\,. 
The inner error bars represent the statistical errors only, the outer 
bars represent the total uncertainty.}
\label{varyxcut}
\end{figure}

Fig.\,\ref{varyxcut} shows the $p_{\perp}$ and $\Delta\phi$ 
distributions together with the ZEUS data. 
One can see that NLO QCD does describe the 
ZEUS data  well without any intrinsic $k_T$ or resummation taken into account.
A reason for the fact that no need for resummation near  $\Delta\phi=\pi$
or $p_{\perp}=0$ is visible  might be that the bin size 
is relatively large. 

Since there is an uncertainty in the reconstruction of 
$x_{\gamma}^{meas}$, we investigated the effect of  shifting the cut 
on $x_{\gamma}^{meas}$, as illustrated in fig.\,\ref{varyxcut}. 
The reason why the result for the last bin   
$p_{\perp}\to 0$ respectively $\Delta\phi\to \pi$ decreases if the 
lower cut on  $x^{meas}_{\gamma}$ is decreased is related to the fact that 
for $x^{meas}_{\gamma}>0.8$ more inelastic processes contribute than for  
$x^{meas}_{\gamma}>0.9$, such that the ratio of nearly back-to-back events to 
the full cross section decreases. 

\begin{figure}[htb]
\begin{center}
\mbox{\epsfig{file=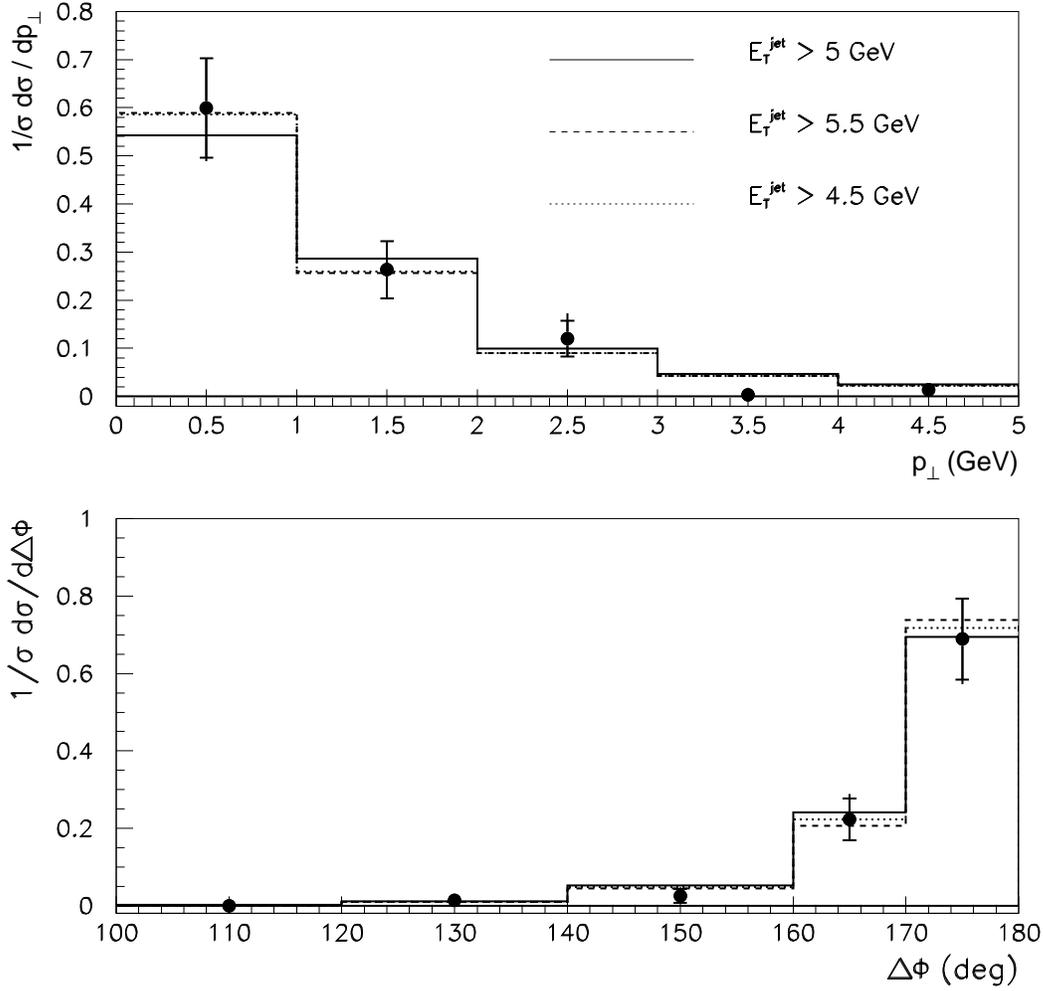,height=15cm}}
\end{center}
\caption{Normalized cross section differential in $p_{\perp}$ and $\Delta\phi$
for  $E_{T,\rm min}^{jet}=4.5, 5$ and 5.5\,GeV\, with $x_{\gamma}^{meas}>
0.9$\,. The inner and outer error bars correspond to statistical 
and total errors, respectively.}
\label{etmin}
\end{figure}

Note that we used symmetric cuts 
$E_{T,\rm{min}}^{jet}=E_{T,\rm{min}}^{\gamma}=5$\,GeV in order to 
match the ZEUS cuts. As discussed above, the theoretical result 
shows instabilities for symmetric cut values. 
Therefore we investigated the effect 
of varying $E_{T,\rm{min}}^{jet}$ by 10\% around the critical point 
$E_{T,\rm{min}}^{\gamma}=5$\,GeV, as shown in fig.\,\ref{etmin}. 
One can see that the result varies by less than 10\% in the 
normalized cross sections. 
The effect of an uncertainty in the cuts on the 
photon energy $y$ has been found to be negligible in the area normalized 
plots. 

Hence we can say that the agreement between fixed order NLO QCD and data
is very good in the case studied here. No $k_T$ in 
addition to the one already implicitly accounted for by including 
the full set of NLO corrections is needed. 
We recall that he same was true for the inclusive 
case~\cite{Fontannaz:2001ek}, but of course there the sensitivity to 
$k_T$ is smaller. 

\medskip

Further we note that the {\it intrinsic} $k_T$  
(i.e. the component of $\langle k_T\rangle$ that does not include the 
parton shower contribution) determined 
by the ZEUS collaboration  
to fit best the data has been obtained by comparing to simulations with 
PYTHIA 6.129. 
On the other hand, it was also found that the data are already well 
described by HERWIG 6.1 with the default parton 
$\langle k_T^{\rm{intr}}\rangle$ 
of zero~\cite{Chekanov:2001aq}.  
The reason might stem from differences in the way the 
parton showering is implemented in  HERWIG and PYTHIA. HERWIG 
does not use a sharp lower cut-off 
in the shower evolution below which PYTHIA relies on a suitable value 
of $\langle k_T^{\rm{intr}}\rangle$.

\clearpage

\section{Conclusions}

We have presented a full next-to-leading order calculation for the 
photoproduction of an isolated photon accompanied by a jet, based on 
a program of partonic event generator type. 

We found that the variation of our result due to scale changes is 
very weak. We also investigated the sensitivity of the cross 
section to  variations of the photon energy, 
to different cuts on $x_{\gamma}$ and
to the cut  $E_{T,\rm{min}}^{jet}$ on the jet transverse energy. 
In particular, we discussed in detail 
the instabilities arising at fixed order NLO QCD if symmetric cuts 
on the photon and jet transverse energies are chosen. 

In comparing the $x_{\gamma}$ distribution to ZEUS data 
we found that the agreement is good except 
near $x_{\gamma}\to 1$. For the photon rapidity distribution 
we found that our  prediction is below the data
in the backward region and overpredicts the data in the forward region. 
We  discussed in detail possible sources of discrepancies
between theory and data. 

\medskip
 
We also compared our results to recent ZEUS data 
for $ k_T $ - sensitive observables. 
Since the comparison has been done 
for normalized cross sections, the effect of variations 
in  the photon energy, $x_{\gamma}^{\rm{cut}}$ or 
$E_{T,\rm{min}}^{jet}$ turned out to be very small. 
We found that NLO QCD already describes these data well without 
introduction of any additional intrinsic $k_T$.

Our results are in contrast to the results found in the E706 fixed target 
experiment~\cite{Apanasevich:1998hm},
where the conclusion was that high-$p_T$ prompt photon and $\pi^0$ data 
could not be well described 
by NLO perturbative QCD without the incorporation of  substantial
extra $k_T$ effects.

\vspace*{8mm}       

{\bf\Large Acknowledgements}

\vspace*{4mm}

We are grateful to P.~Bussey from the ZEUS collaboration for 
helpful discussions. 
G.H. would like to thank the LAPTH for its continuous  hospitality. 
This work was supported by the EU Fourth Training Programme  
''Training and Mobility of Researchers'', network ''Quantum Chromodynamics
and the Deep Structure of Elementary Particles'',
contract FMRX--CT98--0194 (DG 12 - MIHT).


\appendix
\section{Appendix}

In this appendix we identify the origin of the pathological behaviour 
of the cross section in the vicinity of $E_{T,\rm{min}}^{\gamma}$ 
for symmetric cuts on the 
photon and jet transverse energies (see fig.\,\ref{etcut}). 
Our aim is to isolate only the leading behaviour, 
i.e. we work in the double 
logarithmic approximation and drop subleading contributions.

\medskip

We consider the subprocess $p_1+p_2\to p_{\gamma}+p_4+p_5$ where $p_5$ is the 
momentum of a gluon. This gluon represents a NLO correction to the Born
subprocess  $p_1+p_2\to p_{\gamma}+p_4$ and in the soft approximation 
($|\vec{p}_5|\ll |\vec{p}_{\gamma}|$) this NLO correction can be written as
\begin{equation}
\sigma^{NLO}\sim \int d\phi_5 dp_{T5}\;p_{T5}^{-1-2\epsilon}\,
\ln\left(\frac{p_{T\gamma}}{p_{T5}}\right)\;\sigma^{\rm{Born}}(p_1+p_2\to p_{\gamma}+p_4)\;.
\end{equation}
The cut $E_{T,\rm{min}}^{jet}\le E_{T}^{jet} (\equiv p_{T4}$
in the approximation we consider here) restricts the gluon phase 
space to values 
of $p_{T5}$ smaller than $p_{T\gamma}$ and depending on $\phi_5$: 
$p_{T5}^{\rm{max}}=\overline{p}_{T5}(\phi_5)$. 
For $E_{T,\rm{min}}^{jet}\le p_{T\gamma}$  we obtain 
\begin{equation}
\sigma^{NLO}\sim \sigma^{\rm{Born}}\int_0^{2\pi} d\phi_5
\left(\frac{1}{4\epsilon^2} -\frac{1}{2}\ln^2{\frac{p_{T\gamma}}{\overline{p}_{T5}}}
\right)\;.
\end{equation}
The $1/\epsilon^2$ term is cancelled by the virtual contribution. The $\phi_5$ 
integration is not straightforward because the constraint 
$E_{T,\rm{min}}^{jet}\leq p_{T4}$ does not lead to a simple expression of 
$\overline{p}_{T5}(\phi_5)$, but when $E_{T,\rm{min}}^{jet}$ is close to 
$p_{T\gamma}$ we get
\begin{equation}
\sigma^{NLO}\sim \sigma^{\rm{Born}}\times\left(-\frac{1}{2}
\ln^2{\frac{p_{T\gamma}}{p_{T\gamma}-E_{T,\rm{min}}^{jet}}}\right)\;.
\end{equation}
When $E_{T,\rm{min}}^{jet}>p_{T\gamma}$ we obtain 
\begin{equation}
\sigma^{NLO}\sim \sigma^{\rm{Born}}\times\left(\frac{1}{2}
\ln^2{\frac{p_{T\gamma}}{E_{T,\rm{min}}^{jet}-p_{T\gamma}}}\right)\;.
\end{equation}
The cross section displayed in fig.\,\ref{etcut} is integrated over $p_{T\gamma}$
from a lower limit $p_{T,\rm{min}}^{\gamma}$ to $p_{T\gamma}=\sqrt{s}$. 
Therefore the soft NLO contribution to the cross section for 
$E_{T,\rm{min}}^{jet}<p_{T\gamma}$ is
\begin{eqnarray}
-\frac{1}{2} \int_{p_{T,\rm{min}}^{\gamma}}^{\sqrt{s}}dp_{T\gamma}\,
\sigma^{\rm{Born}}(p_{T\gamma})\,
\ln^2{\left(\frac{p_{T\gamma}}{p_{T\gamma}-E_{T,\rm{min}}^{jet}}\right)}\nonumber\\
\approx
-\frac{1}{2}\frac{\overline{\sigma}}{(E_{T,\rm{min}}^{jet})^{n_{eff}-1}}
\int_0^{\frac{E_{T,\rm{min}}^{jet}}{p_{T,\rm{min}}^{\gamma}}}
du\,u^{n_{eff}-2}\ln^2{(1-u)}\label{diverge}
\end{eqnarray}
where we used the approximation 
$\sigma^{\rm{Born}}\approx\overline{\sigma}\,(p_{T\gamma})^{-n_{eff}}\,
(n_{eff}\approx 3)$. 
When $E_{T,\rm{min}}^{jet}$ increases up to $p_{T,\rm{min}}^{\gamma}$, 
the negative contribution (\ref{diverge}) to the cross section 
becomes more and more dominant and explains the behaviour of the cross 
section in the range $E_{T,\rm{min}}^{jet}\leq p_{T,\rm{min}}^{\gamma}$. 
When $E_{T,\rm{min}}^{jet}>p_{T,\rm{min}}^{\gamma}$, the integral over 
$p_{T\gamma}$ splits into two parts
$$\int_{p_{T,\rm{min}}^{\gamma}}^{\sqrt{s}}dp_{T\gamma}=
\int_{p_{T,\rm{min}}^{\gamma}}^{E_{T,\rm{min}}^{jet}}dp_{T\gamma}+
\int_{E_{T,\rm{min}}^{jet}}^{\sqrt{s}}dp_{T\gamma}\;,$$
the second part being calculated already in (\ref{diverge}). 
The first integral is a positive contribution to the NLO cross section, 
given by an expression similar to (\ref{diverge}): 
$$\frac{1}{2}\frac{\overline{\sigma}}{(E_{T,\rm{min}}^{jet})^{n_{eff}-1}}
\int_1^{\frac{E_{T,\rm{min}}^{jet}}{p_{T,\rm{min}}^{\gamma}}}
du\,u^{n_{eff}-2}\ln^2{(u-1)}\;.$$
This positive contribution explains the  behaviour of the cross section 
for $E_{T,\rm{min}}^{jet}>p_{T,\rm{min}}^{\gamma}$.

\end{document}